\documentclass{iau}
\usepackage{graphicx}

\title[JD 6.~~High Energy and Magnetic Processes.... in the Galactic Center
]{Origin of Nonthermal Emission from the Fermi Bubbles and
Mechanisms of Particle Acceleration There}

\author[V. A. Dogiel et al.]   %% give here short author list %%
{V. A. Dogiel$^{1}$\thanks{V.A.Dogiel thanks for  grants from the
IAU  and  RFFI (grant 12-02-00005) that gave him an opportunity to
participate in the IAU303 Symposium.}, K.-S. Cheng$^2$\thanks{KSC
is supported by a GRF grant under HKU 701013.}, D. O.
Chernyshov$^{1}$ \& C.-M. Ko$^{3}$}

\affiliation{ $^1$ P.N.Lebedev Institute of Physics, Leninskii pr.
53, 119991 Moscow, Russia\\ email: {\tt dogiel@lpi.ru,
chernyshov@lpi.ru}\\[\affilskip]$^2$Department of Physics, University of Hong
Kong, Pokfulam Road, Hong Kong, China \\ email: {\tt
hrspksc@hku.hk}\\[\affilskip] $^3$Institute of Astronomy,
National Central University, Chung-Li 32054, Taiwan\\ email: {\tt
cmko@gm.astro.ncu.edu.tw}}

\pubyear{2013}
\volume{303}  %% insert here IAU Symposium No.
\pagerange{119--126}
 \jname{The Galactic Center: IAU 303: Feeding
and Feedback in a Normal Galactic Nucleus} \editors{A.C. Editor,
B.D. Editor \& C.E. Editor, eds.}
\begin{document}

\maketitle

%\begin{abstract}
% \keywords{Keyword1, keyword2, keyword3, etc.}
%\end{abstract}


 %\section{}
 %\vspace*{0.1 cm}
{\it Introduction.} The discovery of the two giant gamma-ray lobes
(Fermi Bubbles) in the Galactic Center (see \cite[Dobler et al.
2010]{dobler10} and \cite[Su et al. 2010]{meng}) was one of the
most impressive events of the last few years in astrophysics.
However,
 some indications on giant structures in the GC were observed
several years before  by WMAP in the radio frequency range between
23 and 33 MHz (\cite[Finkbeiner 2004]{fink}) , and by ROSAT in
hard X-rays (\cite[Bland-Hawthorn \& Cohen 2003]{bland2003}).
Recent observations performed by the Planck Collaboration
(\cite[Ade et al. 2012]{ade12}) found also lobes in the microwave
range which spatially coincided the Fermi Bubbles that indicated
on the common origin of these phenomena.

Parameters of emission from the Fermi bubbles have several
remarkable distinction:
\begin{enumerate}
\item The structures are symmetrically elongated in the direction
perpendicular to the Galactic Plane;
\item Spectra nonthermal emission from the Bubbles  are harder than
anywhere in the Galaxy;
\item The spatial distribution of emission in the Bubbles shows sharp edges of
the Bubbles;
\item The surface emissivity is almost uniform inside the Bubbles although findings of \cite{hooper13}
might indicate on features of the gamma-ray spectrum at latitudes
$b\leq 20^\circ$ which they interpreted as a contribution from the
dark matter annihilation nearby the GC.
\end{enumerate}

The origin of the Bubbles is actively discussed in the literature.
Thus, \cite{crocker11} and \cite{zubovas12}  suggested the
hadronic origin of gamma-ray emission from the Bubbles, when
gamma-ray photons are produced by collisions of relativistic
protons with that of the background gas. Alternatively, these
gamma-rays can be produced by the inverse Compton scattering of
relativistic electrons on background photons (leptonic model) and
the same electrons generate radio and microwave emission from the
Bubbles via synchrotron (see e.g. \cite[Su et al. 2010]{meng}).
There may be several sources (processes) which generate electrons
in the Bubbles:
\begin{itemize}
\item In-situ stochastic acceleration by MHD-turbulence nearby the Bubble
surface (\cite[Mertsch \& Sarkar 2011]{mertsch});
\item Acceleration by shocks which result from periodical star
accretion onto the central black hole (\cite[Cheng et al.
2011]{cheng});
\item Acceleration within jets near the GC about $\sim 10^6$ yr ago, and  subsequent electron transfer
into the bubble by convective flows (\cite[Guo et al. 2012]{} and
\cite[Yang et al. 2013]{zweibel}).
\end{itemize}
Below we discuss some of these models.

{\it Stochastic acceleration from the background plasma.} In order
to reproduce the spatial distribution of gamma-ray emissivity in
the bubble \cite[Mertsch \& Sarkar (2011)]{} assumed arbitrarily
that: a) the acceleration is nonuniformly distributed inside the
Bubbles and its efficiency increases near the shock which excited
the MHD-turbulence inside the bubble; b) the  maximum energy of
 electrons is a function of the distance to the shock.

There are no other evident sources for electrons accelerated in
the halo except electrons from the background plasma and electrons
injected by supernova remnants (SNRs) or jets. Estimates of
acceleration efficiency in the case of stochastic (Fermi)
acceleration from the background plasma is not trivial. As
\cite[Wolfe \& Melia (2006)]{} and \cite[Petrosian \& East
(2008)]{} showed, the energy supplied by sources of stochastic
acceleration is quickly dumped into the thermal plasma because of
ionization/Coulomb energy losses of accelerated particles. As a
result this acceleration is accompanied by plasma overheating
while a tail of nonthermal particles is not formed, i.e. the
effect of stochastic acceleration is negligible.

However, latter \cite[Chernyshov et al. (2012)]{chern12} concluded
that the efficiency of stochastic acceleration depended strongly
on parameters of acceleration and prominent tails of nonthermal
particles can be generated by the acceleration although the
conclusions of \cite[Wolfe \& Melia (2006)]{} and \cite[Petrosian
\& East (2008)]{} are correct for some conditions.

To define whether the stochastic mechanism is able to produce
enough accelerated electrons needed for the observed flux of
gamma-rays we take the following parameters of the background
plasma in the Bubbles:   the  density $n=10^{-2}$ cm$^{-3}$ and
the temperature $T=2$ keV (see \cite[Su et al. 2010]{meng}).

The kinetic equation for the  distribution function electrons,
$f(p,t)$, when processes of spatial propagation are neglected, has
the form
\begin{equation}
 {{\partial f}\over{\partial t}}+{1\over p^2}{\partial\over{\partial
 p}}p^2\left[\left(\frac{dp}{dt}\right)_C f - \left\{D_C+D_F(p)\right\}{{\partial f}\over{\partial
 p}}\right]=0\,,
 \label{eq_nr}
\end{equation}
 $(dp/dt)_C$ and $D_C(p)$ describe particle energy losses and diffusion in the
momentum space due to Coulomb collisions.  The  stochastic (Fermi)
acceleration is  described as diffusion in the momentum space with
the coefficient $D_F(p)$, which we take in the form: $D_F(p) =
\alpha p^\varsigma \theta(p-p_0)$ where $\alpha$, $\varsigma$ and
$p_0$ are arbitrary parameters.

Parameters of this model can be restricted from the three
conditions:
\begin{enumerate}
\item The energy of electrons emitting gamma-rays by inverse
Compton is restricted by the value $\sim 10^{12}$ eV (\cite[Su et
al. 2010]{meng} and \cite[Cheng et al. 2011]{cheng});
\item The total
gamma-ray flux  at energies $E>1$ GeV is $F_\gamma\simeq 4 \times
10^{37}$ erg s$^{-1}$ that restricts the number of accelerated
electrons (\cite[Su et al. 2010]{meng});
%\item The power of potential sources of particles in the GC cannot exceed the
%value about $10^{39}-10^{41}$ erg s$^{-1}$  and \cite[Crocker \& Aharonian 2011]{crocker11});
\item Mechanism  of particle acceleration  should
effectively generate nonthermal particles i.e . no plasma
overheating (\cite[Chernyshov et al. 2012]{chern12}).
\end{enumerate}
\begin{figure}[h]
% \vspace*{-2.0 cm}
\begin{center}
 \includegraphics[width=3.0in]{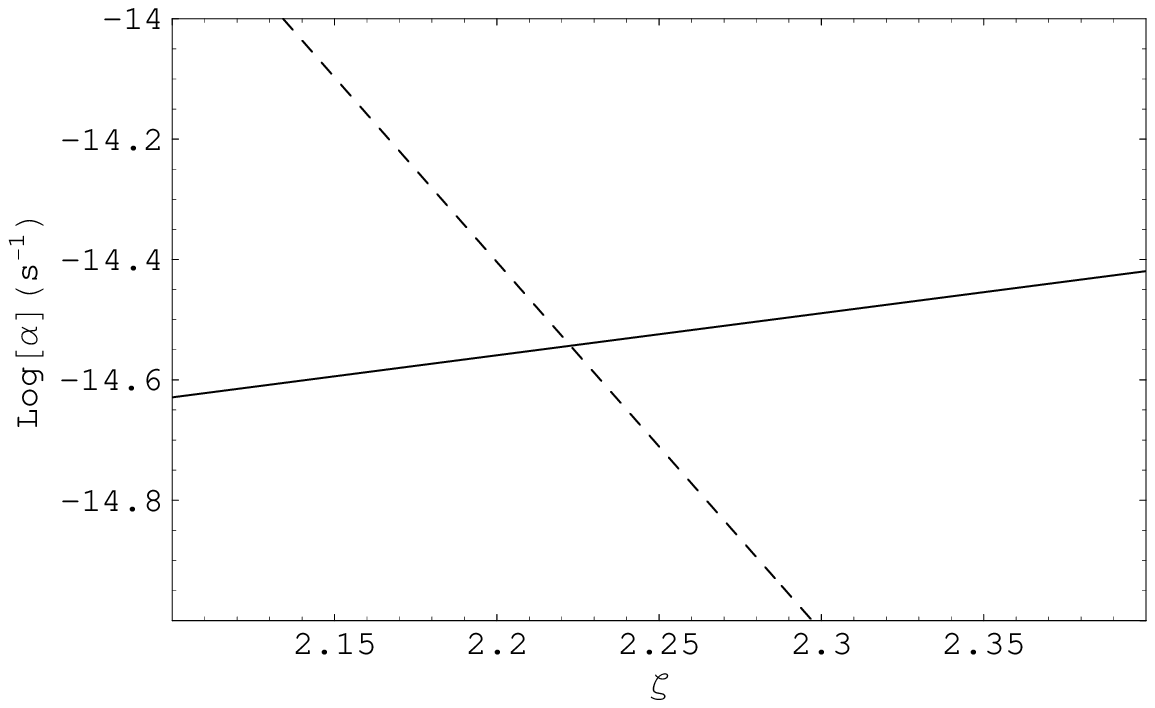}
 \includegraphics[width=2.0in]{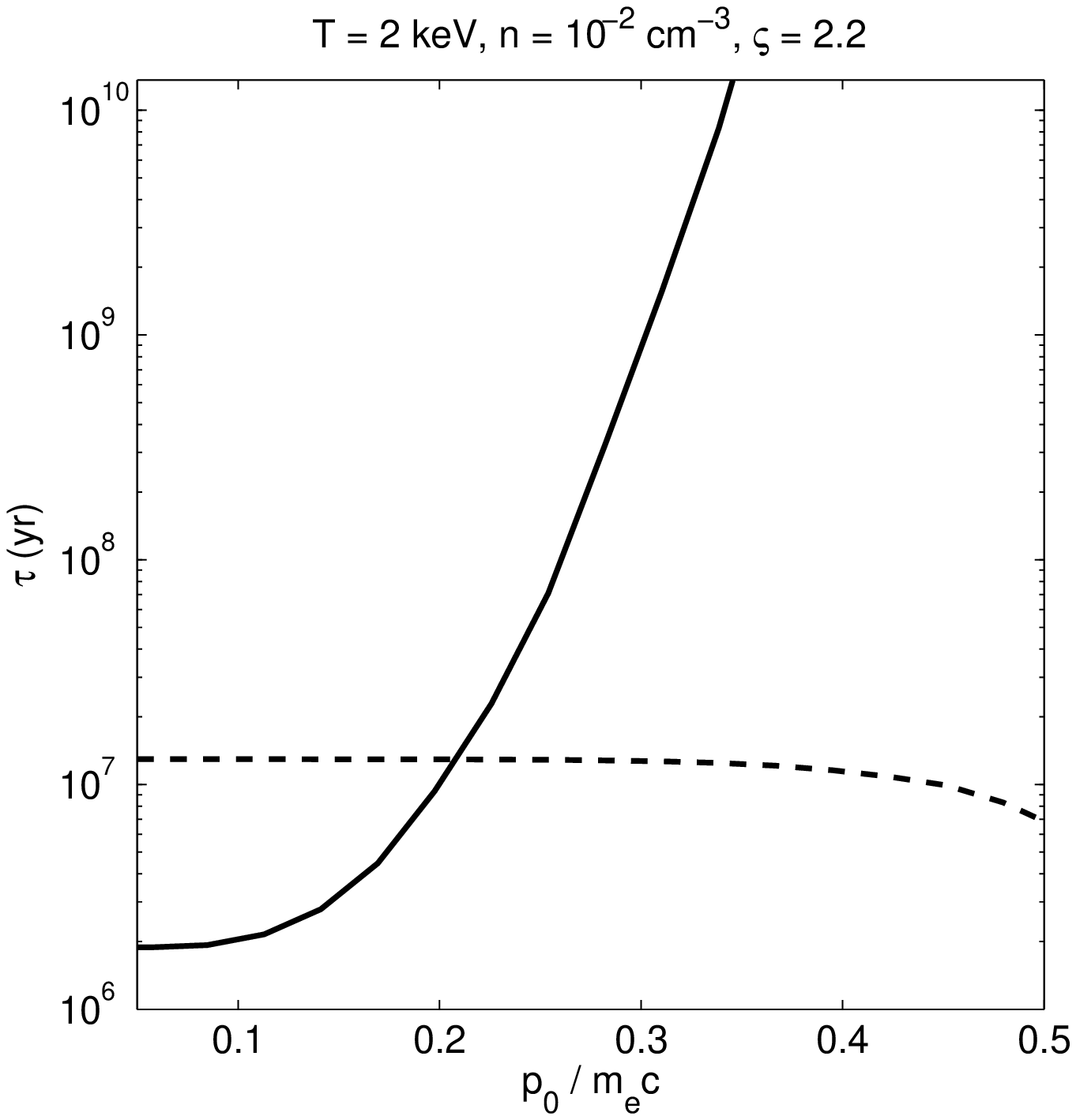}
% \vspace*{-1.0 cm}
 \caption{{\it Left panel:} The functions of $\alpha(\zeta)$ as  derived from the conditions:
 a) -  dashed dotted line; b)  - solid line. {\it  Right panel:} The timescales of temperature variations (solid line) and
acceleration (dashed line) for different values of $p_0$
(condition c).}
   \label{fig1}
\end{center}
\end{figure}
From the results of numerical calculations  shown in Fig. 1 we
conclude that the stochastic acceleration from background plasma
may provide the  density and the spectrum of accelerated electrons
needed for the observed gamma-ray and radio emission from the
Bubbles if the parameters of the model are: $\zeta=2.2$,
$\alpha=3.2\times 10^{-15}$s$^{-1}$, $p_0=(0.2-0.5)mc$. The
required power supplied by sources of acceleration is about
$10^{38}$ erg s$^{-1}$ that does not exceed the upper limit of the
rate of energy release expected in the GC   which is in the range
from $10^{39}$ (\cite[Crocker \& Aharonian 2011]{crocker11}) to
$10^{41}$ erg s$^{-1}$  (\cite[Cheng et al. 2011]{cheng}).

{\it Stochastic re-acceleration of relativistic electrons emitted
by SNRs in the Galactic Disk.} In this case the kinetic equation
has the form
\begin{equation}
-\nabla \left[D(r,z,p)\nabla f -u(r,z)f\right]+
\frac{1}{p^2}\frac{\partial}{\partial p}p^2\left[
\left(\frac{dp}{dt}-\frac{\nabla {\bf u}}{3}p\right)f -
\kappa(r,z,p)\frac{\partial f}{\partial p}\right] =
Q(p,r)\delta(z) \,, \label{eq_nu}
\end{equation}
where $r$ is the galactocentric radius, $z$ is the altitude  above
the Galactic plane, $p$ is the momentum of electrons, $u$ is the
velocity of the Galactic wind, $D$ and $\kappa$ are the spatial
and momentum (stochastic acceleration) diffusion coefficients,
$dp/dt$ describes the rate of electron energy losses, and $Q$
describes the spatial distribution of CR sources in the Galactic
plane ($z=0$) and their injection spectrum.

As it follows from our hydrodynamic numerical simulations the
process of re-acceleration of electrons is supposed to take place
 high above the Galactic plane in regions where the
required MHD-turbulence is excited. Therefore because of the
synchrotron and inverse Compton energy losses only relatively low
energy electrons ejected by SNRs can reach this region. The
thickness of re-acceleration region is defined from the intensity
of the observed gamma-ray and radio emission.

In the simplest case the number of electrons reaching the
re-acceleration region can be calculated in the framework of the
diffusion model of CR propagation (see \cite[Berezinskii et al.
1990]{}) when the convection terms are neglected ($u=0$). For
calculations we used the model parameters from \cite[Ackermann et
al. (2012)]{acker}.

Our numerical calculations show that too many high energy
electrons are produced in the re-acceleration region and, thus,
the condition b) can no be satisfied in the model if the electron
spectrum is formed by the acceleration processes only. Formally we
can assume that processes of particle escape from the acceleration
region are essential enough to make the spectrum steeper and thus
to decrease the number of emitting electrons. Indeed, the momentum
spectrum of accelerated particles is power-law, $f(p)\propto
p^{-\delta}$, with the spectral index $\delta$ equaled
\begin{equation}
\delta = \frac{3}{2}
+\sqrt{\frac{9}{4}+\frac{\tau_{acc}}{\tau_{esc}}}\,,
\end{equation}
where the acceleration time $\tau_{acc} \approx \alpha^{-1}$ and
escape time is $\tau_{esc} \approx \Delta r_b^2/4D$. Here $\Delta
r_b$ is the thickness of re-acceleration region and $D$ is the
spatial diffusion coefficient equaled $D(p) =
{4v_a^2p^2}/(6\kappa(p))$. Here  $v_a$ is the Alfven velocity
which is about 35 km/s in the Galactic halo (see \cite[Ackermann
et al. 2012]{acker}). The numerical calculations showed that the
model reproduces the gamma-ray spectrum if $\alpha\sim 10^{-13}$
s$^{-1}$, $\Delta r_b\sim 10$ pc, and $\delta\sim 4$.  We notice,
however, that the pure diffusion model of CR propagation  has
serious restrictions.  In particular, the effect of convective
transfer (Galactic wind) may be essential in the Galaxy as it
follows from observations (see e.g. \cite[Carretti et al.
2013]{carr13}) as well as from theoretical treatments
(\cite[Bloemen et al. 1993]{bloe} and \cite[Breitschwerdt et al.
2002]{breit}). The influence of the wind might decrease the
density of SNR electrons in the halo significantly that possibly
makes the effect of re-acceleration negligible.

{\it Acceleration by shocks generated by processes of tidal
disruption.} We discussed this model in details in \cite[Cheng et
al. (2011)]{cheng}. In principle, this model describes quite
reasonably the spectra of gamma-ray and radio emission from the
Fermi Bubbles and it explains  the shape of the Bubbles because
shocks propagate in the exponential atmosphere perpendicular to
the galactic plane  (see \cite[Kompaneets 1960]{}). However,
serious simplifications were used for our calculations, e.g. we
used the electron spectrum obtained in a stationary approximation
although the situation of shock propagation in the halo is
essentially non-stationary, we did not take into account shock
evolution in the halo etc. However, we suppose that the shock
model of the Bubbles does not have serious objections up to now.
The energy release from processes of star capture by the central
black hole may release a huge energy up to $10^{54}$ erg. A part
of this energy is transformed into a flux of hard X-ray emission.
Very recently Swift detected to giant X-ray flares in normal
galaxies whose luminosity was about $10^{45}-10^{48}$ erg/s (see
e.g. \cite[Bloom et al. 2011]{}).
 Such a huge flux of hard X-rays from Sgr A* may provide an
 observational effect in the Galactic molecular clouds seen at
 present in the form of "Compton echo" (see \cite[Cramphorn \& Sunyaev 2002]{cramp02}).
Besides, some results of observations  have been already
interpreted as traces of past activity of Sgr A* with a very high
energy release (see \cite[Bland-Hawthornet al. 2013]{} and
\cite[Nakashima et al. 2013)]{} that is in favor of our model.

\end{document}